# Network-theory based modeling of avalanche dynamics in percolative tunnelling networks

Vivek Dey[1], Steffen Kampman[2], Rafael Gutierrez[2], Gianaurelio Cuniberti[2,3], Pavan Nukala[1,*]

[1]Centre for Nano Science and Engineering, Indian Institute of Science, Bengaluru 560012, India,

[2]Institute for Materials Science and Max Bergmann Center of Biomaterials, Technische Universität Dresden, 01062 Dresden, Germany

[3]Dresden Center for Computational Materials Science (DCMS), Technische Universität Dresden, 01062 Dresden, Germany

**Abstract:** Brain-like self-assembled networks can infer and analyze information out of unorganized noisy signals with minimal power consumption. These networks are characterized by spatiotemporal avalanches and their crackling behavior, and their physical models are expected to predict and understand their computational capabilities. Here, we use a network theory-based approach to provide a physical model for percolative tunnelling networks, found in Ag-hBN system, consisting of nodes (atomic clusters) of Ag intercalated in the hBN van der Waals layers. By modeling a single edge plasticity through constitutive electrochemical filament formation, and annihilation through Joule heating, we identify independent parameters that determine the network connectivity. We construct a phase diagram and show that a small region of the parameter space contains signals which are long-range temporally correlated, and only a subset of them contains crackling avalanche dynamics. Physical systems spontaneously self-organize to this region for possibly maximizing the efficiency of information transfer.

## I. INTRODUCTION

Crackling behavior has been observed in diverse physical systems including earthquakes [1], charge density waves [2], magnetic interactions [3], domain interactions in ferroelastic and ferroelectric materials (also known as Barkhausan's noise) [4], second order phase transitions [5], biological neuronal systems [6], and percolative networks [7]. A physical system is said to crackle if it exhibits spatiotemporal avalanches (in electrical, magnetic, or acoustic domains) that follow power-law relations shown in Eqs. (1)-(3) below and a characteristic relationship between the exponents as shown in Eq. (4) [8]:

$$P(S) \sim S^{-\tau} \quad (1)$$

$$P(T) \sim T^{-\alpha} \quad (2)$$

$$\langle S \rangle(T) \sim T^{1/\sigma \upsilon z} \quad (3)$$

$$\frac{1}{\sigma \upsilon z} = \frac{\alpha - 1}{\tau - 1} \quad (4)$$

Here, S and T are the avalanche size and duration, respectively, and P(S) and P(T) are their probability distribution functions. $\langle S(T) \rangle$ is the average avalanche size of duration T.

We note that the brain is a self-assembled complex dynamical network of a large number of interconnected neurons and synapses (~$10^{11}$ neurons and $10^{15}$ synapses in the human brain) [9], which self-organizes to criticality in a state of rest. Universal critical dynamics studied in slices of rat cortex, e.g., are marked by crackling noise with $\tau \sim 1.74$, $\alpha \sim 1.96$, and $1/\sigma \upsilon z \sim 1.22$ [10]. Such a critical state is believed to balance stability and adaptability, a possible necessary trait for efficient information processing [11,12]. Thus, a top-down strategy to emulate brain-like computation in inorganic materials systems is to first look for self-assembled inhomogeneous networks that crackle like the brain.

In this context, predominantly self-assembled percolative networks of Sn-nanoparticles [13] and Ag-PVP nanowires [12] have been investigated. These networks exhibit scale-free as well as small-world properties [14]. In-material signal processing applications such as neuromorphic sensing [15], non-linear information processing [16] and reservoir computing [17,18] have been proposed and demonstrated on these platforms.

Although both show avalanche criticality, they crackle with different sets of avalanche critical exponents ($\tau = 2.05$, $\sigma = 2.66$, and $1/\sigma\upsilon z = 1.55$ for nanoparticle networks; $\tau = 2$, $\sigma = 2.3$, and $1/\sigma\upsilon z = 1.3$ for nanowire networks). In other words, they belong to different universality classes [8].

Recently, two different self-organized critical networks were demonstrated to be hosted in compact Ag-hBN (CVD) memristive devices. Two terminal MIM devices comprising Ag-hBN-Cu show non-volatile memristive hysteresis, which can be stabilized either in a high resistance state (HRS) or a low resistance state (LRS) when no stimulus is applied. The LRS is structurally characterized as a network of multiple Ag filaments, whereas the HRS is a network of intercalated Ag clusters within the hBN matrix, see in Fig. 1(a) and Ref. [8]. In both of these states, the device exhibits spatiotemporal avalanches that crackle, resulting from an underlying self-assembled network. However, the crackling noise in HRS is reminiscent of tunneling nanoparticle networks (in terms of critical exponents and universality classes), whereas in the LRS, it resembles nanowire percolative networks.

Since these networks hosted in single memristive devices are self-assembled and plastic, controllability of every node cannot be achieved, unlike in bottom-up assembled memristive crossbars. Any form of computation using these black box networks will require designing efficient learning algorithms with limited control nodes (multiple electrode platforms, e.g.). In this context, it is imperative to first model and understand the physics of the "blackbox network" between two controllable nodes (source and ground electrodes) on which external stimulus can be applied.

The current study introduces a network theory-based modeling approach focused on the underlying network behavior of Ag-hBN-Cu device set to HRS. Every edge in the network shows plasticity driven by electromigration of Ag, joule heating, and quantum mechanical tunneling transport. The investigation identifies key independent parameters influencing network connectivity and establishes a phase diagram utilizing these connectivity parameters. We show that crackling behavior exists only in a limited parameter space of this phase diagram, identifying the self-tuning required in the network to exhibit self-organized criticality (SOC).

## II. PHYSICAL NETWORK GENERATION

STEM-EDS data on the HRS of the Ag-hBN-Cu memristor device shows that it can be described as a network of intercalated Ag clusters/nodes (see in Fig. 1(a)). The background conductivity of this state arises from quantum mechanical tunneling between the nearest nodes. We first generate our physical network in an L x L-sized box (L~10-20 nm of physical dimension) based on these experimental results. The box is divided into $l$ layers, with separation $d_0$ representing interlayer spacing in h-BN. The nodes (representing Ag atomic clusters) are randomly distributed in each layer. An edge is assigned between two neighboring nodes if these nodes are within maximum tunneling separation, $D_{max}$, set by a minimum off-conductivity value $G_{min}$:

$$G_{ij} = G_0 \exp(-\beta D) \tag{5}$$

We set $G_{min}$ to a very low value of $10^{-15}$ $\Omega^{-1}$; $G_0$, or maximum possible tunnel conductance ($D\rightarrow 0$) as 1 $\Omega^{-1}$; and β, the inverse tunneling decay length as 100 nm$^{-1}$, as justified in Ref. [16]. At $D=0$, when a full filament is formed between the nodes (physical connection, referred to as an edge filament), we assign $G_{ij\text{-on}} = G_0 = 100$ $\Omega^{-1}$. It may be noted that the exact values of network conductance ($G_{on}$ especially) are not important; rather, the trends that dictate fluctuations in the conductance are.

Figure 1(b) shows the visualization of a connected network with nodes and edges (between neighbors), as defined above. Note that voltage (external stimulus) can be applied only to the source node (red in Fig 1(b)), representing the Ag top electrode. The ground node (at ground potential always) represents the Cu back electrode (green node in Fig. 1(b)).

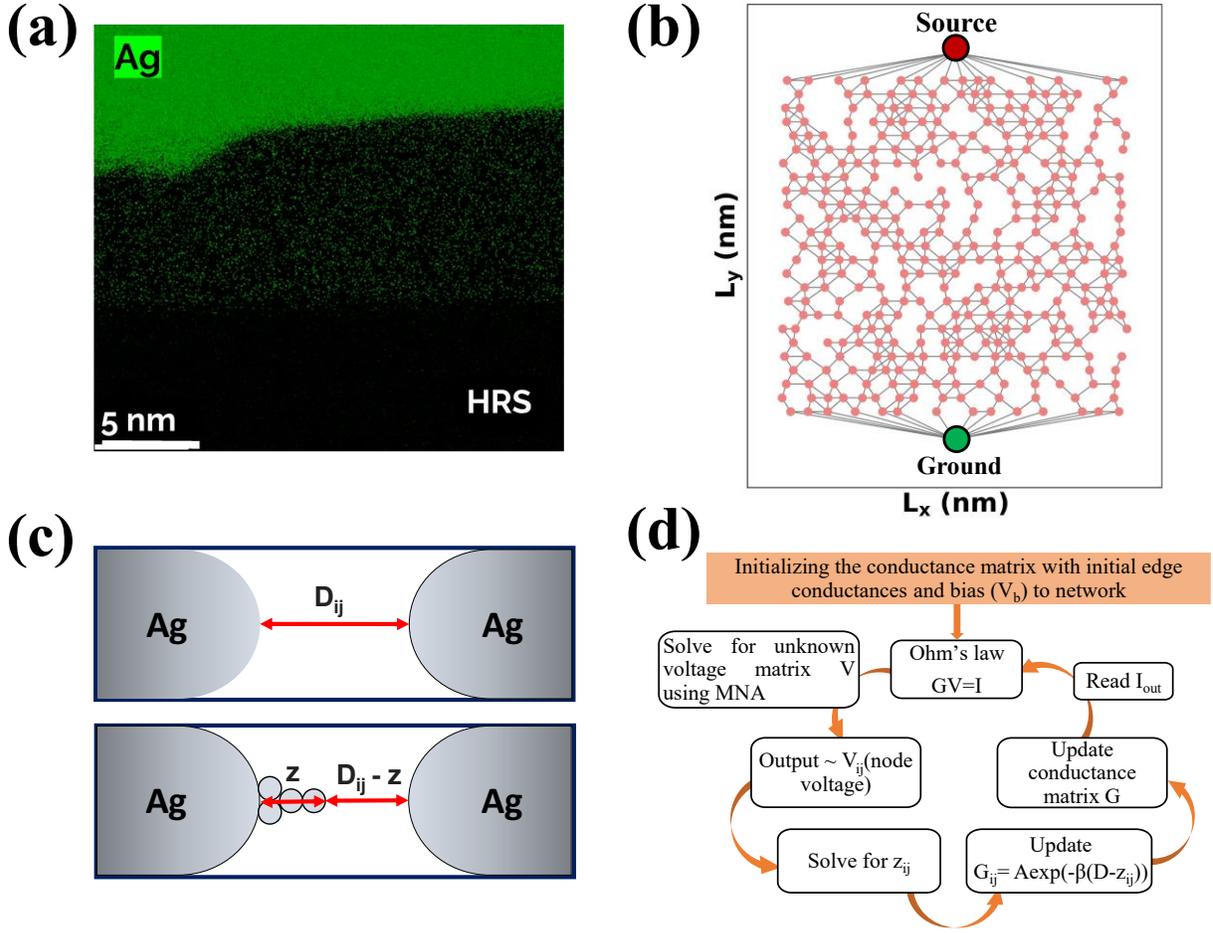

FIG. 1. **Ag-cluster network and edge model**. a) STEM-EDS image of Ag/hBN/Cu (MIM) device. Bias is applied to the Ag electrode, and the Cu electrode is always grounded. b) Graphical representation of the as-generated connected network of clusters with sourcecode (red) and ground node (green). c) Schematic of the edge modeling of the electromigration of $Ag^+$ and effective change in the cluster separation. d) A diagram representing the computational steps involved in one simulation cycle (MNA, modified nodal analysis).

## III. NETWORK EDGE DYNAMICS MODEL

Dynamics at every edge arise owing to the following factors:

a) Formation of an atomic wire through $Ag^+$ electromigration that reduces the tunnel gap, increasing the edge conductivity.

b) Dissolution of the edges via Joule heating when the edge current increases beyond a threshold value, reducing the edge conductivity.

The former process (formation of an atomic wire) is modeled by a modified mobility equation [17]:

$$\frac{dz_{ij}}{dt} = \frac{\mu V_{ij}}{D_{ij}-z_{ij}} - kz_{ij}, \tag{6}$$

with the first term on the right-hand side representing atomic wire growth and the second term representing its decay in the absence of voltage due to thermodynamic instability. Here, $z_{ij}$ is the length of the atomic wire, $V_{ij}$ is the edge voltage difference that drives the formation of the atomic wire, and $D_{ij}$ is the effective distance between $i^{th}$ and $j^{th}$ clusters, as shown in Fig 1(c). $\mu$ is a parameter governing the growth of the atomic wire, and $k$ is the rate constant for the decay process. The values of $\mu$ and $k$ are taken to be 0.346 nm² V⁻¹ and 0.038 s⁻¹, chosen for the atomic wire's formation and decay time scale to be similar [17].

We rewrite Eq. 6 in dimensionless form as follows:

$$\frac{dz'_{ij}}{dt'} = \frac{\mu V'_{ij}}{kD_{ij}^2(1-z'_{ij})} - z'_{ij} \tag{7}$$

where, $z'_{ij} = z_{ij}/D_{ij}$ and $t' = kt$. Eq. (3) can be solved recursively using Euler's method with a simulation timestep $\Delta t > 0$ as:

$$z'_{ij,t} = z'_{ij,t-1}(1 - k\Delta t) + k\Delta t \left[\frac{\mu V_{ij}}{kD_{ij}^2(1-z'_{ij,t-1})}\right] \tag{8}$$

Here, $z'_{ij,t}$ and $z'_{ij,t-1}$ are normalized atomic wire lengths at time t and t-1, respectively, between clusters i and j. In the above equation, $V_{ij}$ is calculated using Kirchhoff's current law at every simulation timestep (see computational flow chart in Fig. 1(d)).

Next, we model the effects of Joule heating by introducing two additional parameters (i) a conductance cut-off, $G_{cut}$, and (ii) a reduction factor, RF. $G_{cut}$ (in units of $G_{on}$) is the maximal conductance in an edge, above which the edge filament (EF) dissolves because of excessive heating. The parameter RF models the multiple steps EF dissolution process. When RF=1, an EF of size 1 will completely dissolve once $G \geq G_{cut}$. However, when RF > 1, only a fraction of the EF (of the order 1/RF) will dissolve in one time step when $G \geq G_{cut}$. RF also determines the rate of EF dissolution, that is shown in Supplemental Material [19]. We further assume that the effect of Joule heating kicks in only after the edge conductance ($G_{ij}$) exceeds a threshold value of $G_{joule}$. In other words when $G_{ij} \leq G_{joule}$, EF dynamics is controlled only by Eq. 8. All in all, the complete Joule heating induced filament dissolution process is described by:

$$z'_{ij,t} = z'_{ij,t-1} - \frac{1}{RF} \; ; \quad if \; (G_{cut} - G_{joule}) \leq G \leq G_{cut} \tag{9}$$

The algorithm employed to model the full dynamics is shown in Fig. 2.

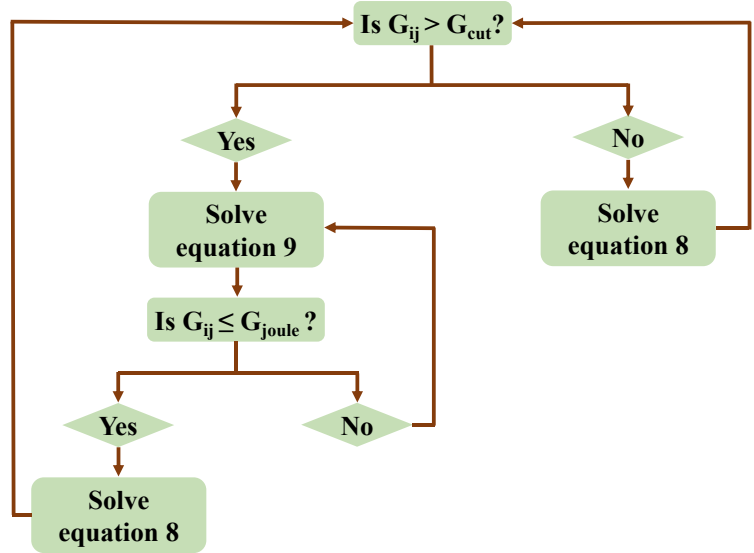

FIG. 2. **Edge-filament dissolution algorithm**. The flowchart shows the algorithm used to describe the overall dissolution process of the edge filament.

## IV.   RESULTS AND DISCUSSION

Using the algorithm displayed in Fig. 2, we simulated the network dynamics for 200 secs starting from the initialized state when biased at $V_b$ = 1V. The current response is recorded over every simulation timestep for different combinations of $G_{cut}$ and RF (Figs. 3(a) and 3(b)), showing different kinds of bursting events, which will be analyzed subsequently. At larger $G_{cut}$, we see that the background network conductivity gradually increases with time (Fig. 3(b)), a result which is intuitively clear.

Next, we show the physical evolution of the network with time at two particular values of $G_{cut}$ and RF. At $G_{cut}$=0.01$G_{on}$ and RF=10, we see multiple edge filaments forming and breaking (consecutive red circled region in top panel of Fig 3(c), also see Supplemental Video 1 in Supplemental Material [19]), emulating the HRS characteristics of Ag-hBN memristor, and corresponding current fluctuations [8]. At $G_{cut}$=0.1$G_{on}$, RF=10 (Fig 3(c) bottom panel, also see Supplemental Video 2 in Supplemental Material [19]) we see the evolution of multiple physical filaments from the source to the node. This emulates the HRS to LRS first order phase transition in Ag-hBN memristors presented in Ref. [8] (also shown in Supplemental Material [19]).

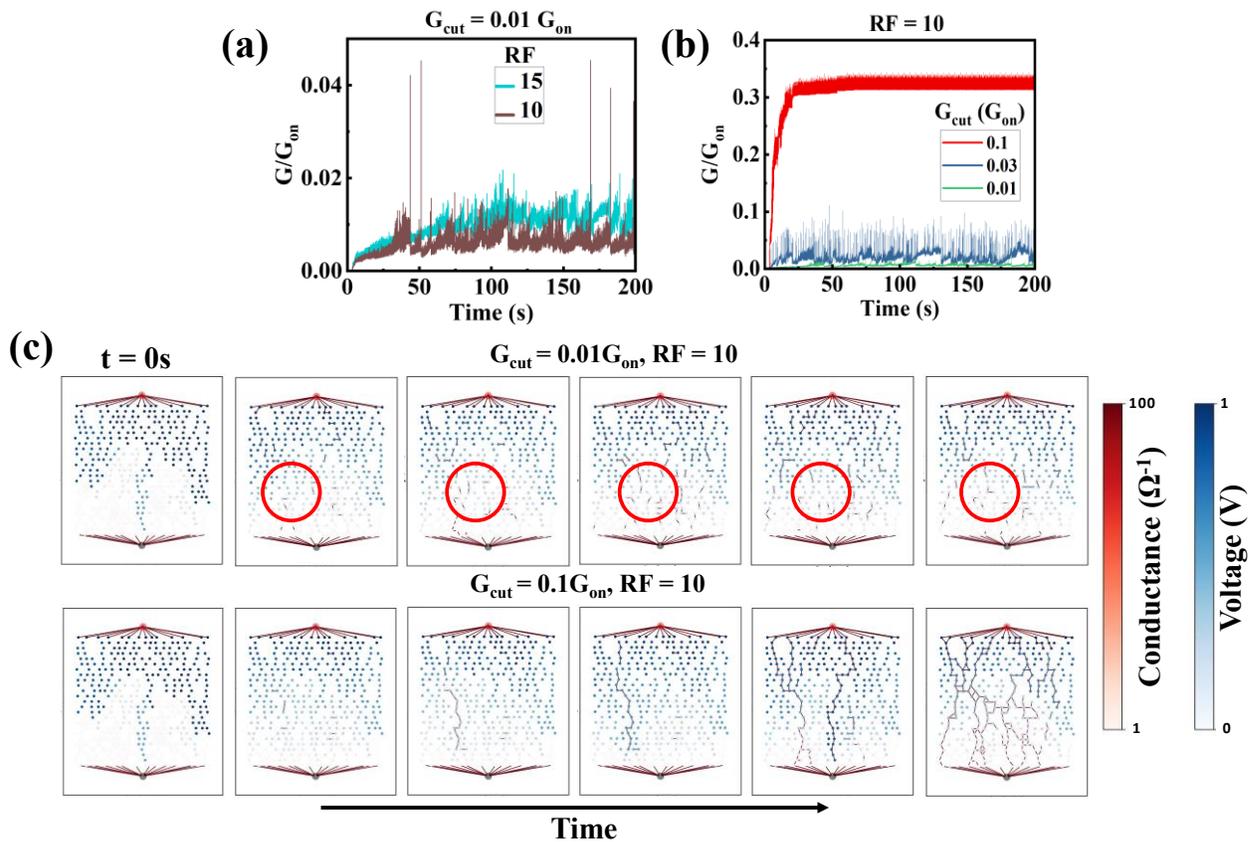

FIG. 3. **Network evolution**. Simulated current vs. time data for (a) $G_{cut}$ = 0.01 and different RF, (b) different $G_{cut}$ and RF = 10, (c) Top panel: graphical representation of the time evolution of the edge filaments at $G_{cut}$=0.01$G_{on}$ and RF=10, showing multiple EF forming and dissolving with time in red circles representing the HRS and bottom panel: time evolution of the edge filaments at $G_{cut}$=0.1$G_{on}$ and RF=10, multiple filaments evolve (high conductance percolating path) from source to ground representing LRS of the Ag/hBN/Cu device.

The power spectrum at $G_{cut} = 0.01G_{on}$ and RF=10, follows a power law scaling with frequency as $1/f^{\beta}$ with exponent $\beta = 0.8$ over three decades. However, at $G_{cut} = 10^{-4} G_{on}$ and RF = 5, the network does not show such 1/f noise over such interval (see Supplemental Material [19]).

We note that the combination of $G_{cut}$ and RF values determine the network connectivity (NC), and the behavior of order parameters associated with it. Here, we show the time series evolution of order parameters $\Delta G(G_0)$ (conductance change) and IEI (Inter event interval), and their respective PDFs for two different network connectivity combinations (Figs. 4(a)-4(f)). At $[G_{cut}, RF]=[10^{-4} G_{on}, 5]$ (referred to NC-A), the PDFs of both $\Delta G$ and IEI (Figs. 4(b) and (c)) decay faster than a power law. Whereas for $[G_{cut}, RF]=[0.01G_{on}, 10]$ (referred to as NC-B), these parameters show a power law distribution, with well defined exponents (Figs. 4(e) and (f)). We also see that the autocorrelation function $A(t)$ of IEIs in NC-B displays a weak power law spanning over more than 4 temporal decades, demonstrating long range temporal correlations (LRTC) between signals. However, in NC-A, $A(t)$ is smaller by three orders of magnitude, suggesting negligible LRTCs (Fig. 4(g)).

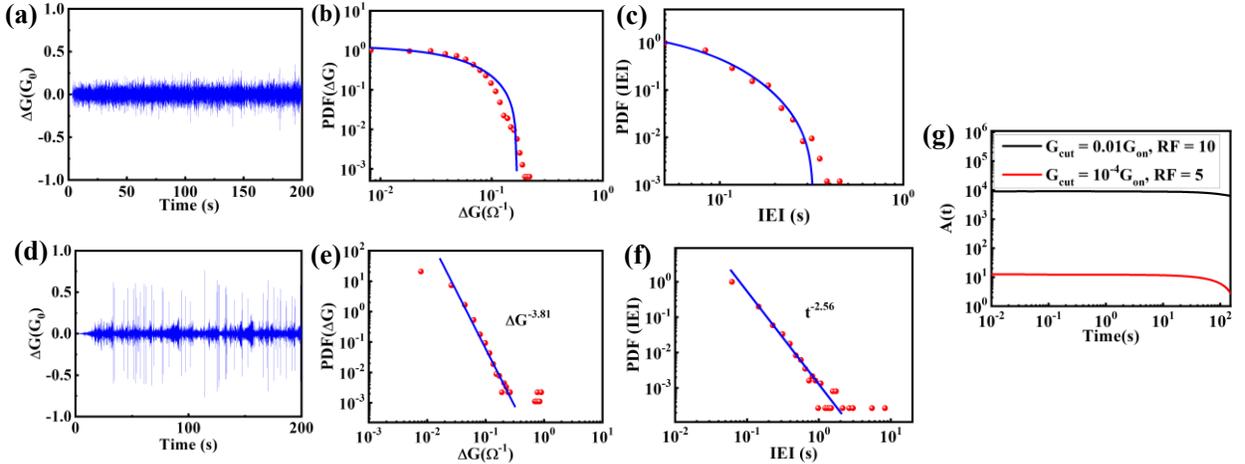

FIG. 4. **Long-range temporal correlation and power law scaling.** (a) and (d) are the difference in conductance representing the event sizes $\Delta G$ (blue), expressed in quanta of mean conductance, $G_0$, for NC-A and NC-B respectively. (b) and (c) The probability distribution (PDF) plots of $\Delta G$ and inter-event intervals (IEI) for network connectivity NC-A. (e) and (f) are PDF of $\Delta G$ and IEI respectively for network connectivity NC-B. (g) Autocorrelation function of IEIs for both NC-A and NC-B. The sampling rate for simulation is 10ms.

To understand which of these networks crackle, next, we analyze the avalanche statistics in both NC-A and NC-B networks (Eqs. 1-3). In NC-A, the PDF of avalanche size (S) and duration (T) also decays faster than a power law within less than two decades (Figs. 5(a) and 5(b)), and thus shows no crackling. However, in NC-B, S, T and <S(T)> follow power laws over more than two decades (Figs. 5(d)- 5(f)), with well-defined exponents, τ (1.73), α (1.51) and 1/σvz (1.11). In fact, these exponents also follow the relation shown in Eq. 4, demonstrating that NC-B is a network that shows crackling noise, similar to the experimental results shown in the HRS state of Ag-hBN system [8]. The time series data analysis protocols used here are similar to those used in Ref. [8].

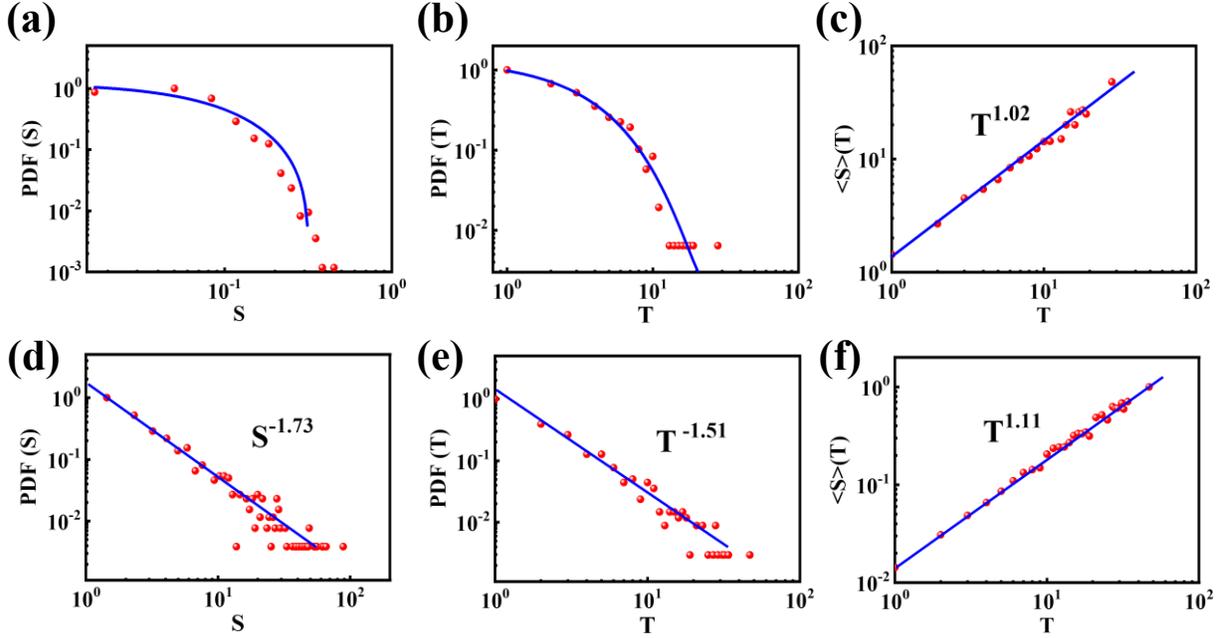

FIG. 5. **Crackling noise**. The distribution plots (a) to (c) are for NC-A and (d) to (f) are for NC-B. (a) and (d) The probability distribution function (PDF) for size of avalanche $P(S)$ follows exponential distribution for the NC-A and power law for the NC-B with exponent values of $\tau \sim$ 1.73. (b) and (e) PDF of avalanche duration $P(T)$ follows exponential distribution for NC-A whereas it follows power law for NC-B with exponent value of $\alpha \sim 1.51$ with linear binning. (c) and (f) The average avalanche size per unit time bin $<S>(T)$ both follows power law with exponents $1/\sigma vz \sim 1.02$ (NC-A) and $1/\sigma vz \sim 1.11$ (NC-B).

Finally, we construct a phase diagram to illustrate what network connectivity parameters (NCP), $G_{cut}$ and RF, make a network crackle (Fig. 6). In Fig. 6 we show the behavior of order parameter IEI at various NCPs, with color scale representing the decades over which power law scaling is valid. The purple color represents all the NCPs where the IEIs decay faster than a power law, and do not exhibit long range temporal correlations. We repeat this exercise (of constructing phase diagrams) for $S$, $T$ and $<S(T)>$ and show that only a subset of NCPs (dashed region) in Fig. 6 that shows IEI power law distribution (in yellow), exhibits crackling noise or avalanche criticality. Interestingly, the network always self-organizes, experimentally, into this small region of NCPs.

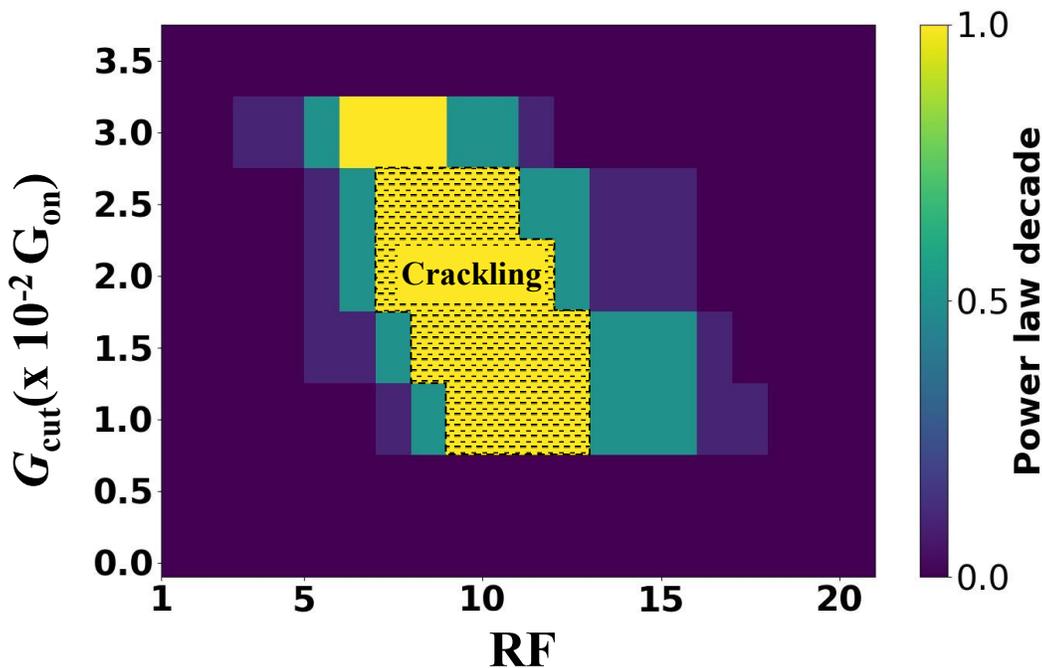

FIG. 6. **Map of parameter space**. Parameter phase map for combinations of network connectivity parameters $G_{cut}$ and RF. The regions are highlighted as per the power law decade of order parameter IEI. The network shows crackling behavior in the shaded yellow region of the parameter map.

V. CONCLUSION

We have presented a network theory-based model to emulate the stochastic device dynamics of Ag-hBN memristors set to HRS. Our network consists of Ag nodes intercalated randomly between hBN layers, with network current dynamics dictated by physical processes at every edge, such as electromigration of Ag, joule heating, and internodal quantum mechanical tunneling. Through this model, we identified key parameters governing network connectivity (NCP) and established a phase diagram relating these parameters to characterize the network activity and avalanche dynamics. We have shown that avalanche criticality and crackling happen only in a very tight region of the phase space of NCPs. We posit that physical systems such as neural networks self-tune themselves to operate in this region, possibly to transfer information efficiently, given the long-range temporal correlations. Our physical modeling of the network provides the first step towards solving computational problems using these systems, from intrinsically unstructured and noisy data, similar to what biological neural networks do.

**Acknowledgement:** The authors acknowledge the DAAD travel grant between IISc and TU Dresden, that enabled the collaboration and the work. P.N. acknowledges Start-up grant from IISc, Infosys Young Researcher award, and SERB (DST), New Delhi, Govt. of India CRG/2022/003506. PN and VD acknowledge the usage of national nanofabrication center, micro nano characterization center, and advanced facility for microscopy and microanalysis of IISc for various fabrication and characterization studies. PN and VD acknowledge discussions with Ankit Rao from University of Minesotta, USA and Nithin Nagaraj from NIAS, Bengaluru, India.

**REFERENCES:**

1. A. Sornette and D. Sornette, *Self-Organized Criticality and Earthquakes*, Europhysics Letters **9**, 197 (1989).
2. C. R. Myers and J. P. Sethna, *Collective Dynamics in a Model of Sliding Charge-Density Waves. I. Critical Behavior*, Phys Rev B **47**, 11171 (1993).
3. B. Casals, G. F. Nataf, and E. K. H. Salje, *Avalanche Criticality during Ferroelectric/Ferroelastic Switching*, Nature Communications **12**, 1 (2021).
4. P. J. Cote and L. V. Meisel, *Self-Organized Criticality and the Barkhausen Effect*, Phys Rev Lett **67**, 1334 (1991).
5. J. P. Sethna, K. A. Dahmen, and C. R. Myers, *Crackling Noise*, Nature **410**, 242 (2001).
6. N. Friedman, S. Ito, B. A. W. Brinkman, M. Shimono, R. E. L. Deville, K. A.Dahmen,J.M.Beggs, and T. C. Butler, *Universal Critical Dynamics in High Resolution Neuronal Avalanche Data*, Phys Rev Lett **108**, 208102 (2012).
7. E. K. H. Salje and X. Jiang, *Crackling Noise and Avalanches in Minerals*, Physics and Chemistry of Minerals **48**, 1 (2021).

8. A. Rao, S. Sanjay, V. Dey, M. Ahmadi, P. Yadav, A. Venugopalrao, N. Bhat, B. Kooi, S. Raghavan, and P. Nukala, *Realizing Avalanche Criticality in Neuromorphic Networks on a 2D HBN Platform*, Mater Horiz **10**, 5235 (2023).
9. S. Herculano-Houzel, *The Human Brain in Numbers: A Linearly Scaled-up Primate Brain*, Front Hum Neurosci **3**, (2009).
10. E. Bullmore and O. Sporns, *The Economy of Brain Network Organization*, Nature Reviews Neuroscience **13**, 336 (2012).
11. W. L. Shew, H. Yang, T. Petermann, R. Roy, and D. Plenz, *Neuronal Avalanches Imply Maximum Dynamic Range in Cortical Networks at Criticality*, Journal of Neuroscience **29**, 15595 (2009).
12. [J. Hochstetter, R. Zhu, A. Loeffler, A. Diaz-Alvarez, T. Nakayama, and Z. Kuncic, *Avalanches and Edge-of-Chaos Learning in Neuromorphic Nanowire Networks*, Nature Communications **12**, 1 (2021).
13. S. Shirai, S. K. Acharya, S. K. Bose, J. B. Mallinson, E. Galli, M. D. Pike, M. D. Arnold, and S. A. Brown, *Long-Range Temporal Correlations in Scale-Free Neuromorphic Networks*, Network Neuroscience **4**, 432 (2020).
14. A. Loeffler, R. Zhu, J. Hochstetter, M. Li, K. Fu, A. Diaz-Alvarez, T. Nakayama,J. M. Shine, and Z. Kuncic, *Topological Properties of Neuromorphic Nanowire Networks*, Front Neurosci **14**, 516308 (2020).
15. Z. Ma, W. Ch en, X. Cao, S. Diao, Z. Liu, J. Ge, and S. Pan, *Criticality and Neuromorphic Sensing in a Single Memristor*, Nano Letters **23**, 5902 (2023).
16. E. Baek et al., *Intrinsic Plasticity of Silicon Nanowire Neurotransistors for Dynamic Memory and Learning Functions*, Nature Electronics **3**, 398 (2020).
17. J. B. Mallinson, Z. E. Heywood, R. K. Daniels, M. D. Arnold, P. J. Bones, and S. A. Brown, *Reservoir Computing Using Networks of Memristors: Effects of Topology and Heterogeneity*, Nanoscale **15**, 9663 (2023).
18. R. K. Daniels, J. B. Mallinson, Z. E. Heywood, P. J. Bones, M. D. Arnold, and S. A. Brown, *Reservoir Computing with 3D Nanowire Networks*, Neural Networks **154**, 122 (2022).
19. See Supplemental Material for further details of the experimental and simulated results, as well as the videos of the simulation.

# Supplementary Information

# Network-theory based modeling of avalanche dynamics in percolative tunnelling networks

*Vivek Dey[1], Steffen Kampman[2], Rafael Gutierrez[2], Gianaurelio Cuniberti[2,3], Pavan Nukala[1]*

[1]Centre for Nano Science and Engineering, Indian Institute of Science, Bengaluru 560012, India,

[2]Institute for Materials Science and Max Bergmann Center of Biomaterials, Technische Universität Dresden, 01062 Dresden, Germany

[3]Dresden Center for Computational Materials Science (DCMS), Technische Universität Dresden, 01062 Dresden, Germany

This file includes:

Fig S1 to S3

VideoS1 and VideoS2

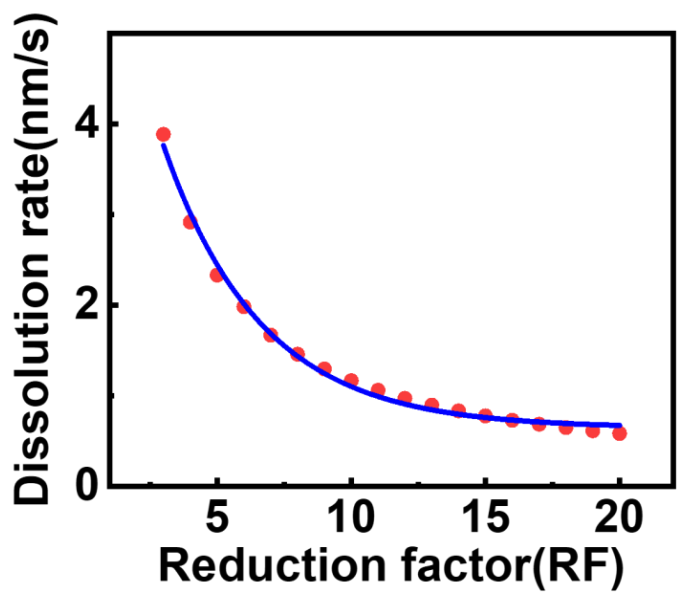

Fig S1: Rate of dissolution of edge filament vs reduction factor (RF).

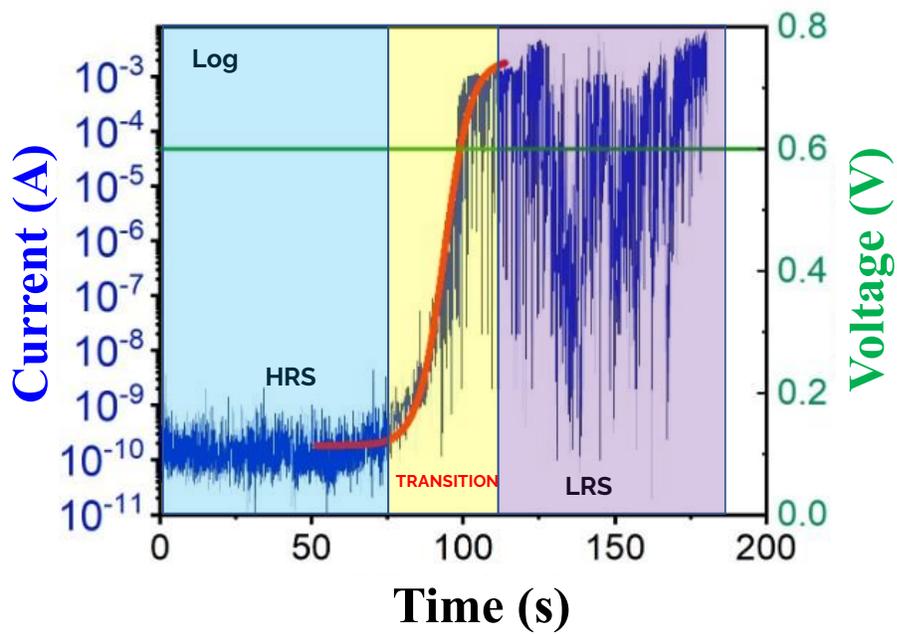

FigS2: HRS to LRS transition in Ag-hBN memristor device [1].

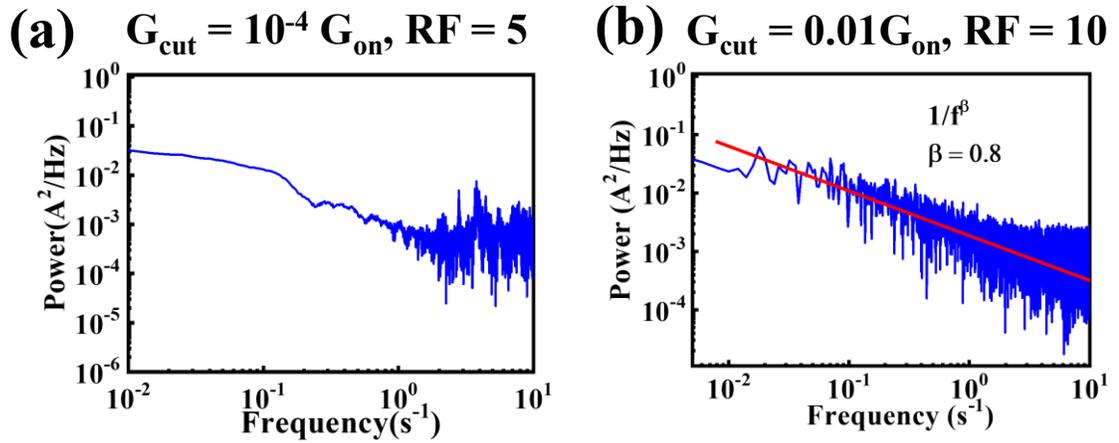

Fig S3: Power spectrum. (A) At $G_{cut} = 10^{-4}G_{on}$ and RF = 10, the power spectrum follows power law with frequency as 1/f. (B) At $G_{cut}=0.01G_{on}$ and RF=5 the power spectrum.

**VideoS1:**

The video shows the local formation and dissolution of edge filaments at $G_{cut}=0.01G_{on}$ and RF=10.

**VideoS2:**

The video shows the time evolution of edge filaments at $G_{cut}=0.1G_{on}$ and RF=10 leading to multiple filament formation.

Link to videos: https://drive.google.com/drive/folders/1gvA9-Qd9xzvD6wfy5sLU4jiLWYcY1Egt?usp=sharing

**References**

[1] A. Rao, S. Sanjay, V. Dey, M. Ahmadi, P. Yadav, A. Venugopalrao, N. Bhat, B. Kooi, S. Raghavan, and P. Nukala, *Realizing Avalanche Criticality in Neuromorphic Networks on a 2D HBN Platform*, Mater Horiz **10**, 5235 (2023).